\PassOptionsToPackage{inner=2cm, outer=3cm, marginparwidth=2cm, marginparsep=0.6cm}{geometry}
\documentclass[3p,a4paper,review]{elsarticle}

\usepackage{marginnote}

\usepackage{footnote}

\usepackage{graphics}
\usepackage{lineno}
\usepackage{subcaption}	

\usepackage{amsmath,amssymb,amsfonts,amsthm,amsbsy,amstext,stmaryrd}
\usepackage{siunitx}		
\usepackage{xfrac}
\usepackage[super]{nth}

\usepackage{booktabs} 	
\usepackage{multirow} 	

\usepackage[usenames,dvipsnames,table]{xcolor}

\usepackage{url}

\newcommand{\no}{\text{---}}

\biboptions{square,comma,sort&compress}

\journal{Journal of Alloys and Compounds}

\begin{document}

\begin{frontmatter}

\title{Solubility and partitioning of impurities in Be alloys.}

\author[UNSW]{P.A. Burr\corref{corresponding}}
\ead{p.burr@unsw.edu.au}
\author[Westinghouse]{S.C. Middleburgh}
\author[IC]{R.W. Grimes}
\address[UNSW]{School of Electrical Engineering and Telecommunications, University of New South Wales, Kensington, 2052, Australia.}
\address[Westinghouse]{Westinghouse Electric Sweden AP, 72163 V\"aster\r{a}s, Sweden.}
\address[IC]{Centre for Nuclear Engineering and Department of Materials, Imperial College London, London, SW7 2AZ, UK.}
\cortext[corresponding]{Corresponding author.}

\begin{abstract}
The most energetically favourable accommodation processes for common impurities and alloying elements in Be metal and Be-Fe-Al intermetallics were investigated using atomic scale simulations.
Fe additions, combined with suitable heat treatments, may scavange Al and Si through their incorporation into the FeBe$_5$ intermetallic. In the absence of Fe, Al and Si will not be associated with Be metal.
Li and Mg are also not soluble, but may react with other impurities if present (such as Al or H). Mg may also form the MgBe$_{13}$ intermetallic phase under certain conditions.
He and H exhibit negligible solubility in all phases investigated and whilst He will tend to form bubbles, H can precipitate as BeH$_2$. 
Similarly, C additions will form the stable compound Be$_2$C.
Finally, oxygen exhibits a strong affinity to Be, exhibiting both some degree of solubility in all phases considered here (though especially metallic Be) and a highly favourable energy of formation for BeO.
\end{abstract}

\end{frontmatter}



\section{Introduction}
\label{sec:intro}

Beryllium (Be) metal is a technologically important material due to its light weight, high stiffness, thermal stability and radiation transparency, and thus sees use in various aerospace applications. Those properties combined with the low atomic number and the remarkable neutronic characteristics, make it an ideal candidate for fusion and fission technologies. It is currently used as a neutron reflector in most water cooled nuclear reactors, as a plasma facing material in the JET fusion reactor \cite{Deksnis1997} and is integral in the design of the ITER fusion reactor  \cite{Thompson2007}.

Be metal often comes with small amounts of impurities that exhibit limited solid solubility. Such impurities are expected to diffuse to surfaces and grain boundaries (thereby worsening the mechanical and chemical properties of the alloy), if they are not retained within the grains by sinks such as second phase particles, point defects, or voids.

\begin{figure}[hb]
\begin{subfigure}[b]{0.32\columnwidth}
                \centering
                \includegraphics[width=\textwidth]{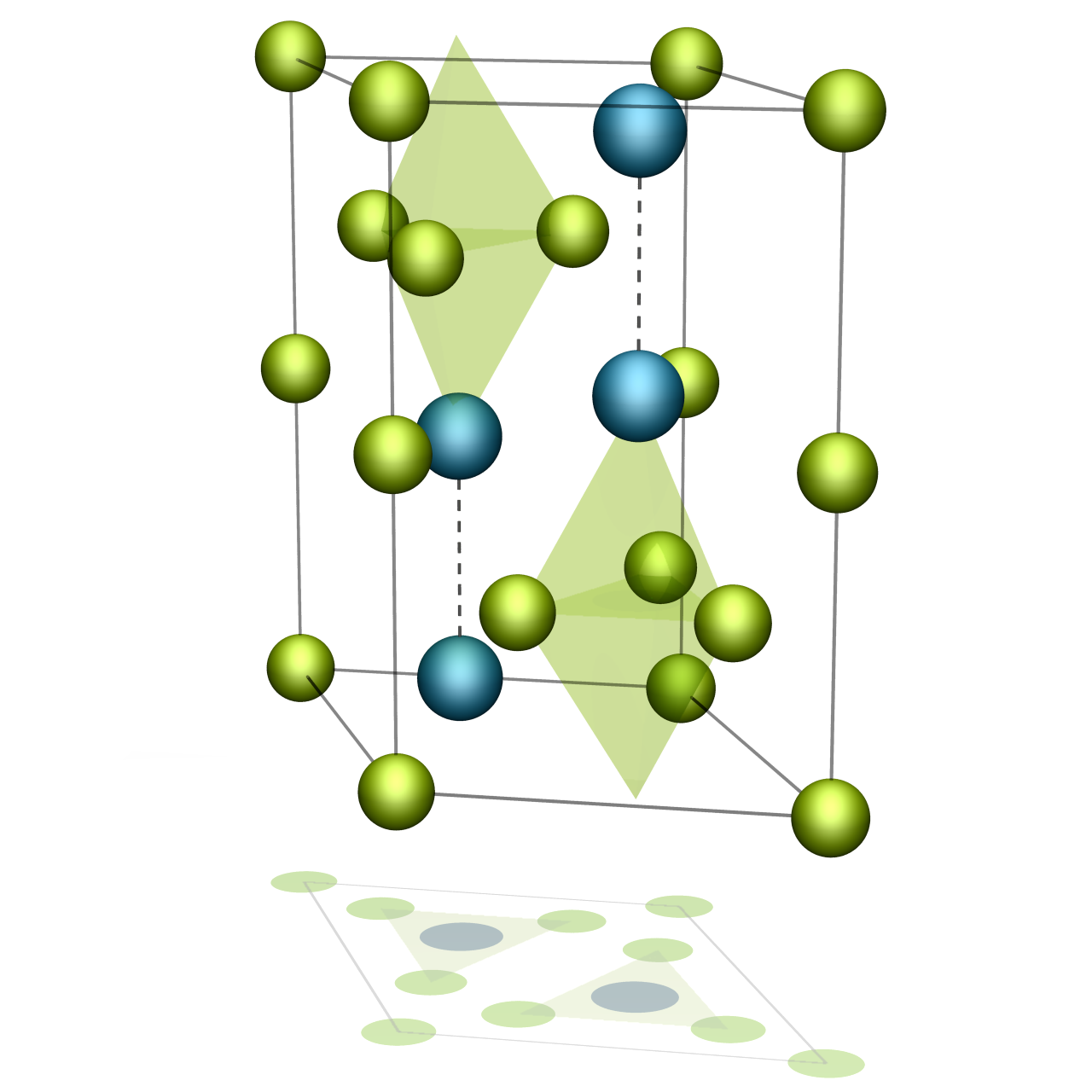}
                \caption{FeBe$_2$}
                \label{fig:FeBe2}
\end{subfigure}
\begin{subfigure}[b]{0.32\columnwidth}
                \centering
                \includegraphics[width=\textwidth]{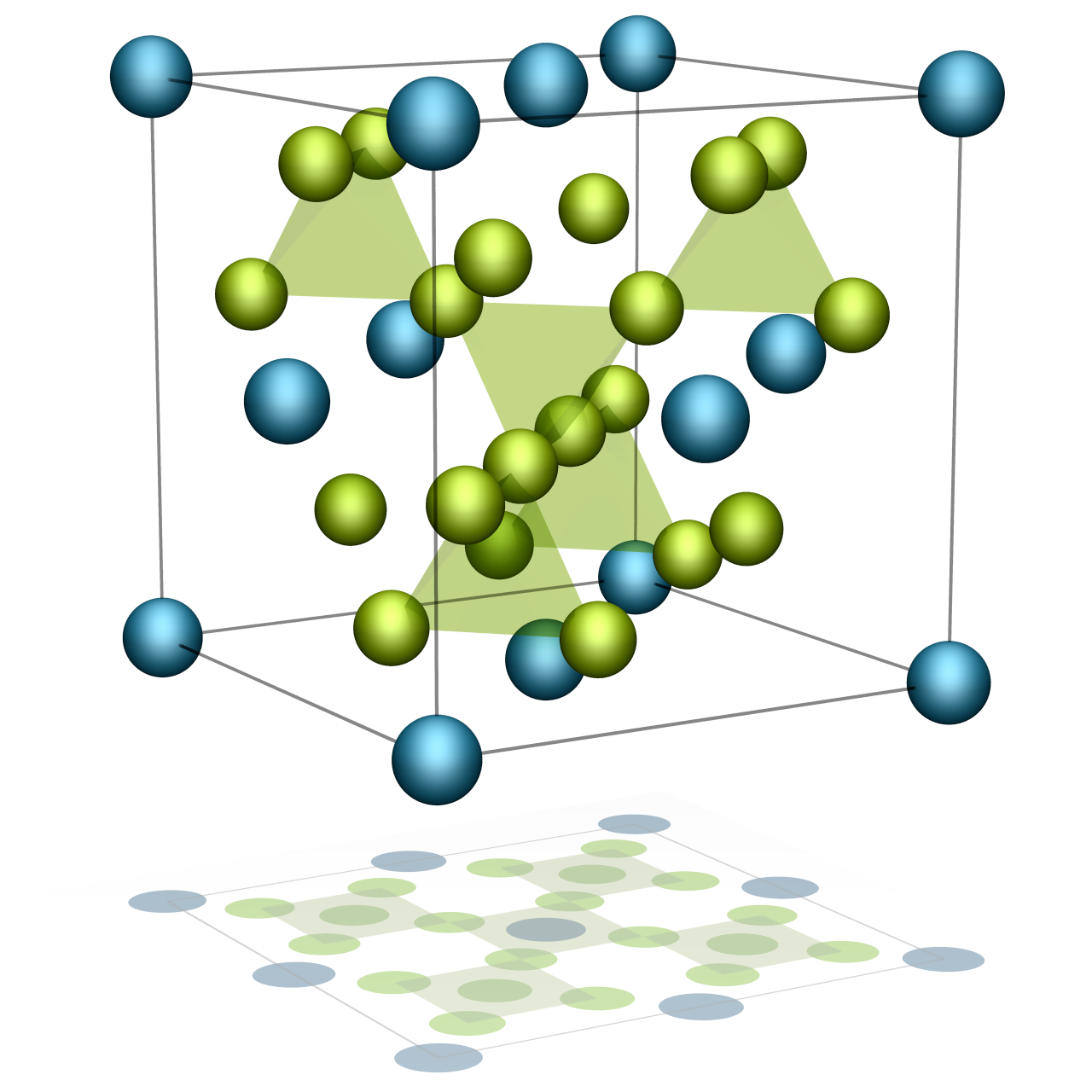}
                \caption{FeBe$_5$}
                \label{fig:FeBe5}
\end{subfigure}
\begin{subfigure}[b]{0.32\columnwidth}
                \centering
                \includegraphics[width=\textwidth]{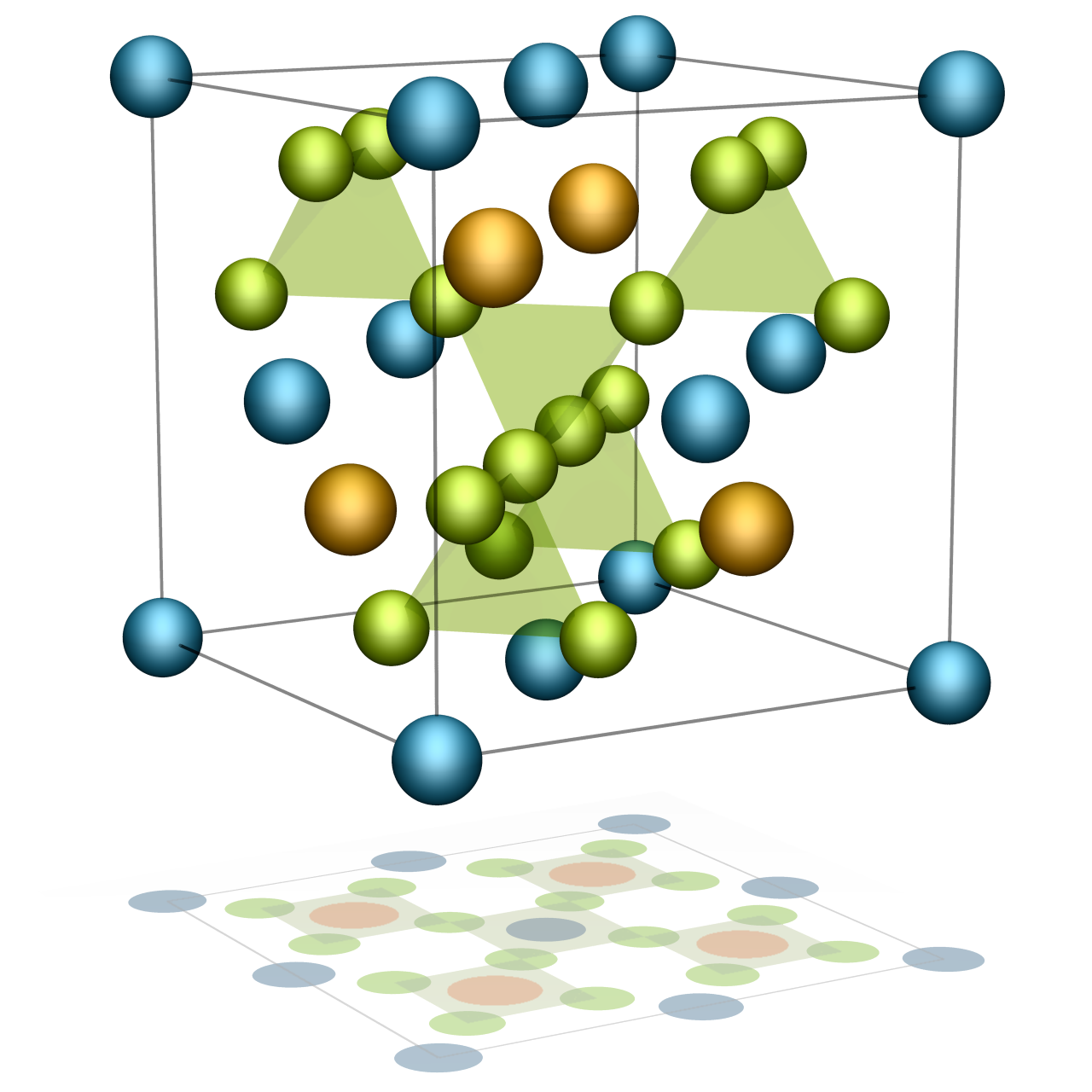}
                \caption{AlFeBe$_4$}
                \label{fig:AlFeBe4}
\end{subfigure}
\caption{Crystal structures of Be intermetallics. Green, blue and orange spheres represent Be, Fe and Al atoms, respectively. Shaded tetrahedra illustrate the local arrangement of Be atoms.}
\label{fig:xtals}
\end{figure}

In a previous publication \cite{Middleburgh2011} the solubility of common impurities --- such as Al, Fe, C, H, He, Li, Mg, O, Si --- in HCP-Be was investigated by means of density functional theory (DFT) based atomic scale simulations.
In a recent publication \cite{Burr2015a}, we show that FeBe$_5$ and AlFeBe$_4$ second phase particles are expected to form in Be alloys containing Fe and Al, and that FeBe$_2$ is expected to form if sufficient Fe is present but no Al. The structures of these intermetallics are depicted in Figure~\ref{fig:xtals}. In the current work we examine the ability of these intermetallics to scavenge other common impurities and compare this behaviour to  solubility limits in HCP-Be.

\section{Computational Methodology}
\label{sec:meth}
The density functional theory (DFT) simulations used in this work employed the {\sc castep} code \cite{Clark2005} using the PBE exchange-correlation functional \cite{Perdew1996}, ultra-soft pseudo potentials and a consistent plane-wave cut-off of \SI{550}{eV}.

For point defect calculations, supercells containing 171 -- 216 atoms were used with a high {\bf k}-point density \cite{Monkhorst1976} (the distance between sampling points was maintained as close as possible to \SI{0.30}{nm^{-1}} and never above \SI{0.35}{nm^{-1}}). In practice this means a sampling grid of $3\times3\times3$ points for the largest supercells. 

Since these systems are metallic, density mixing and Methfessel-Paxton \cite{Methfessel1989} cold smearing of bands were employed with a width of \SI{0.1}{eV}. Testing was carried out to ensure a convergence of \SI{e-3}{eV/atom} with respect to all parameters.
No symmetry operations were enforced when calculating point defects and all calculations were spin polarised, taking particular care that defective cells remained in the same magnetic configuration as the perfect cell.
Point defect simulations were relaxed until the energy difference between two consecutive geometries was less than \SI{1e-6}{\electronvolt}.

In our previous work \cite{Burr2015a}, we showed that lattice disorder plays an important role for AlFeBe$_4$ and --- at high temperatures --- for FeBe$_5$. Calculating the energy associated with point defects within disordered phases is computationally impractical, hence the current work assumed only the ordered form of AlFeBe$_4$ and FeBe$_5$.

\section{Results}
\label{sec:res}

First, the solubility of extrinsic elements in bulk Be was considered and the energy of solution of element $M$ on site $j$ was calculated following equation~\ref{eq:Esol}:
\begin{equation} \label{eq:Esol}
E_{sol} = E^{\text{DFT}}(M_j) - E^{\text{DFT}}(host) \pm \sum_j \mu_j
\end{equation}
where $E^{\text{DFT}}(host)$ and $E^{\text{DFT}}(M)$ are the total energies of the Be supercell before and after the introduction of the defect and $\mu_j$ is the chemical potential of the species that has been removed/added from site $j$ to retain mass action. $\mu$ is calculated via DFT simulations of the elements  in their ground state structure: HCP Be, FCC Al, ferromagnetic BCC Fe, R3-graphite, H$_2$ gas, He gas, low temperature Li ($R3\bar{m}H$), HCP Mg, O$_2$ gas, and crystalline Si with diamond structure.
Only the lowest energy sites, as previously identified~\cite{Allouche2010a,Middleburgh2011,Zhang2012a,Bakai2011}, were re-calculated in the current work and are reported in Table~\ref{tab:extrinsic}. However, since no prior work considered Li accommodation in metallic Be, all potential accommodation sites were simulated. These simulaitons indicate that the most favourable mechanism for Li accommodation is via substitution with a solution energy of \SI{1.01}{eV}. Interstitial solution is less favourable: \SI{6.19}{eV}, \SI{6.11}{eV} and \SI{5.61}{eV}  for octahedral, hexahedral and trigonal interstitial, respectively. Interestingly, when a Li atom occupies a tetrahedral site, it spontaneously relaxes onto a Be site and displaces the Be atom into a crowdion-like defect. This complex defect yields a lower solution energy than any simple Li interstitial (\SI{5.24}{eV}), but still consistent with interstitial Li in HCP-Be, under equilibrium conditions, being highly unfavourable.



\begin{figure}[p]
\centering
\includegraphics[width=4in]{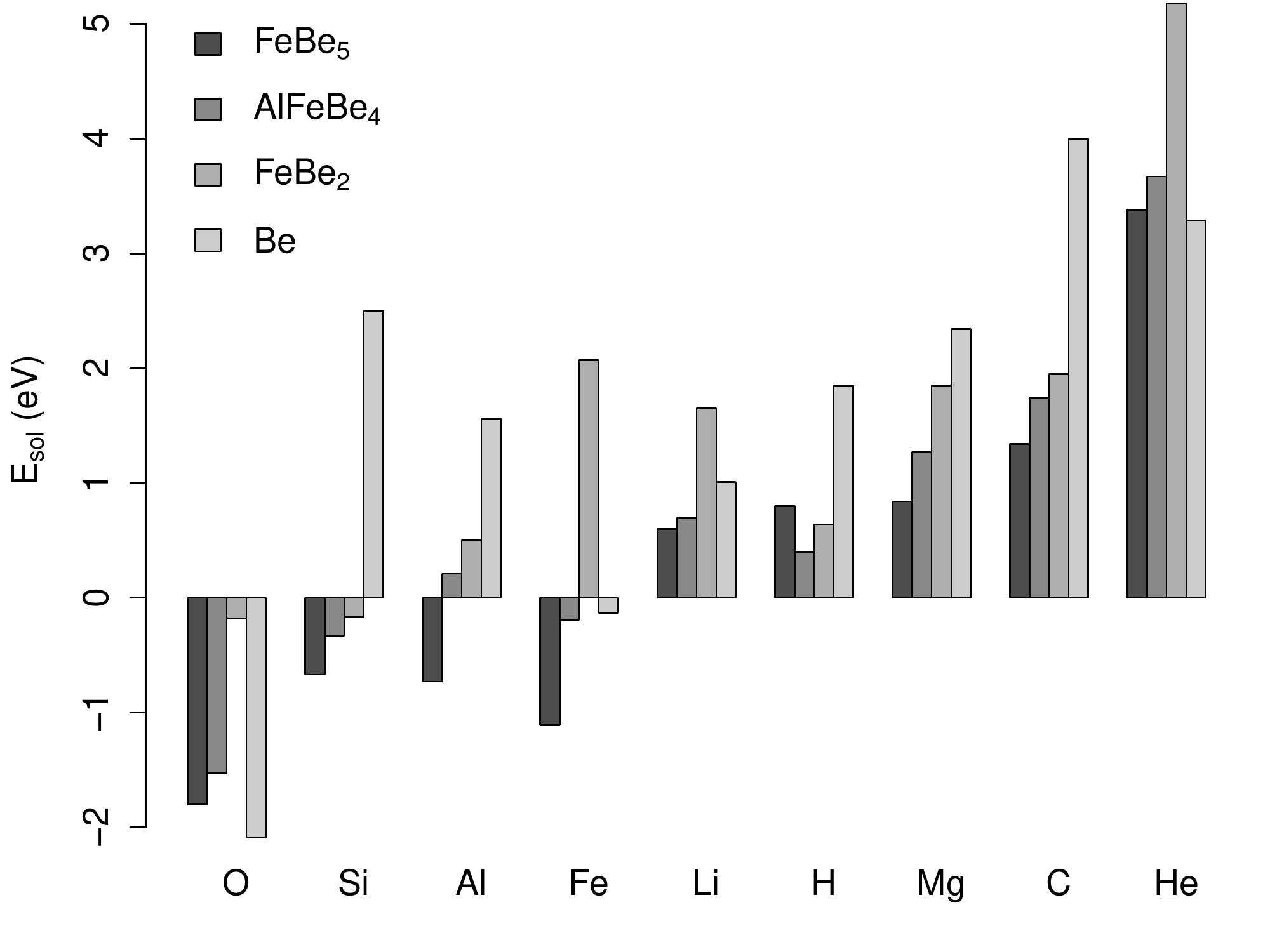}
\caption{\label{fig:extrinsic} Most favourable or least unfavourable solution energies of extrinsic elements in FeBe$_5$, AlFeBe$_4$, FeBe$_2$, Be and Be$_2$C.}
\end{figure}

The accommodation enthalpies of common impurities in intermetallic phases of Be were calculated and compared to the accommodation enthalpies of the same impurities in bulk Be, see Table~\ref{tab:extrinsic}.
Extrinsic species are usually accommodated on the crystal lattice sites via a substitutional mechanism, though, species with relatively small atomic radii or those that may bond covalently, such as C, H and O, may be accommodated at an interstitial site. A dash indicates that interstitial accommodation was found to be considerably less favourable (a few eV greater) compared to substitutional accommodation. For the largest metallic atoms, substitution onto a Be-tetrahedron (e.g.\ Al$_{\{4\text{Be}\}}$) was also considered, but these were found to be consistently less favourable than conventional substitution mechanisms.
The most favourable (or least unfavourable) solution energies for all element/phase combinations are summarised graphically in Figure~\ref{fig:extrinsic}.



\begin{table}[p]
\scriptsize
\centering
\caption{\label{tab:extrinsic}
Solution energy ($E_{\text{sol}}$, in eV) of impurities in FeBe$_2$, FeBe$_5$, AlFeBe$_4$ and Be. Only the lowest energy interstitial configurations are presented. For each extrinsic element, the most favourable accommodation mechanism, is highlighted in \bf{bold}.}
\begin{tabular}{l l S S S S S S S S S S}
\toprule
Phase	&site	 &\text{Al}	&\text{Fe}	&\text{C}	&\text{H}	&\text{He}	&\text{Li}	&\text{Mg}	&\text{O}	&\text{Si}\\
\midrule
\multirow{4}{*}{FeBe$_2$}
&Be$(2a)$	&0.79	&1.07	&2.09	&2.43	&5.18	&1.65	&3.21	&0.46	&-0.14	\\
&Be$(6h)$	&0.95	&1.30	&1.95	&2.27	&5.39	&1.88	&3.25	&0.61	&-0.17	\\
&Fe$(4f)$		&0.50	&\no		&4.97	&3.59	&5.25	&1.66	&1.85	&0.58	&0.58	\\
&$i$			&\no		&5.03	&2.78	&0.64	&5.19	&\no		&\no		&-0.18	&\no	\\
\midrule
\multirow{4}{*}{FeBe$_5$}
&Be$(4c)$	&\bf{$-$0.73}&-1.11	&4.72	&3.68	&4.47	&0.85	&1.05	&1.63	&\bf{$-$0.67}	\\
&Be$(16e)$	&0.94 	&0.63	&1.34	&2.21	&5.00	&1.92	&3.20	&-0.47	&-0.08	\\
&Fe$(4a)$	&-0.22 	&\no		&3.60	&0.81	&3.38	&\bf{0.60}	&0.84	&-0.18	&-0.11	\\
&$i$			&\no	 	&\no		&1.37	&0.80	&4.28	&\no		&\no		&-1.80	&\no		\\
\midrule
\multirow{4}{*}{AlFeBe$_4$}
&Al$(4c)$		&\no		&-0.19	&5.59	&4.45	&4.94	&1.40	&1.40	&2.40	&-0.03	\\
&Be$(16e)$	&0.81	&0.58	&1.74	&2.15	&4.70	&1.54	&3.03	&-0.50	&-0.33	\\
&Fe$(4a)$	&0.21	&\no		&3.85	&3.05	&3.67	&0.70	&1.27	&0.05	&2.45	\\
&$i$			&\no		&4.77	&2.39	&0.40	&4.20	&\no		&\no		&-1.53	&\no		\\
\midrule
\multicolumn{2}{l}{Be(s)} &1.56	&-0.13	&4.00	&1.85	&\bf{3.29}	&1.01	&2.34	&-2.09	&2.50	\\
\midrule
\multicolumn{2}{l}{\multirow{2}{1.5cm}{Intermetallic formation}} 	&
						\multirow{2}{*}{\no}	&
						\text{FeBe$_2$}	&
						\text{Be$_2$C}		&
						\text{BeH$_2$}		&
						\multirow{2}{*}{\no}	&
						\multirow{2}{*}{\no}	&
						\text{MgBe$_{13}$}	&
						\text{BeO}			&
						\text{Be$_2$Si}		\\
&			&	&\bf{$-$1.30}	&\bf{$-$0.68}	&\bf{$-$0.17}	&	&	&\bf{0.06}	&\bf{$-$6.06}	&0.70	\\

\bottomrule
\end{tabular}
\end{table}

\section{Discussion}
\label{sec:disc}

\subsection{Iron and aluminium}
As discussed in previous work~\cite{Burr2015a}, Fe is soluble in Be metal  (i.e. a small negative solution energy), but also forms (more favourably) three intermetallic phases: $\varepsilon$-Fe$_{2+x}$Be$_{17-x}$, $\delta$-FeBe$_{5}$ and $\zeta$-FeBe$_2$, with the latter two commonly observed in commercial alloys. FeBe$_5$ will also accommodate excess Fe leading to a degree of non-stoichiometry.
Al has negligible solubility in Be metal, but is strongly accommodated in $\delta$-FeBe$_{5}$ though not in the other Fe-Be intermetallics. If sufficient Al is present in the system, disordered (Al,Fe)Be$_2$ is formed.

\subsection{Lithium and magnesium}
As stated in section~\ref{sec:res}, the solution energy of Li in Be, \SI{1.01}{eV}, is consistent with very limited solubility.
Of the phases considered, Li is most soluble in FeBe$_5$ and AlFeBe$_4$, but even then the solubility limit will be very small.
No binary Li-Be intermetallics are known to exists (although it has been speculated that an intermetallic may form under extreme pressures \cite{Galav2013}), therefore, Li is available to combine with other minor constituents such as Mg and Al (with which it forms intermetallics \cite{Goel1990,Taylor2010,Okamoto2012b}) or H and O (with which it combines with high heats of formation \cite{CRC90}).

Solution energies for Mg follow a similar trend to those for Li, though they are even less favourable in all cases. 
Conversely, a small positive energy for the formation of the intermetallic phase MgBe$_{13}$ is identified.  It should be noted that the magnitude of this energy is sufficiently small that second order energy contributions not included in the current work (such as thermal vibrations, zero point energy and configurational disorder) may provide sufficient additional contributions to stabilise MgBe$_{13}$. Nevertheless, MgBe$_{13}$ formation remains the lowest energy process to accommodate Mg in Be.

\subsection{Carbon and silicon}
Carbon appears to be highly insoluble in Be metal, in line with previous DFT studies \cite{Middleburgh2011,Zhang2012a}. Carbon solubility in Fe-Al-Be binary and ternary intermetallics is also very limited (solution energy of \SIrange{1.34}{1.95}{eV}), with the least unfavourable accommodation mechanism being substitution for tetrahedrally coordinated Be atoms. C will, however, readily form the compound Be$_2$C following:
\begin{equation}
2 \text{Be}_{(s)} + \text{C}_\text{(graphite)}  \xrightarrow{\SI{-0.68}{eV}}  \text{Be}_2\text{C}_{(s)}
\end{equation}

Si is in the same elemental group as C, yet it exhibits a distinctly different behaviour with respect to Be alloys.
First, the existence of a Be$_2$Si phase, with a Be$_2$C-like structure, while proposed by recent DFT work~\cite{Yan2011}, has not been observed experimentally~\cite{Rooksby1962a}. Our results predict an unfavourable formation energy of \SI{0.7}{eV} for Be$_2$Si, consistent with the lack of experimental observations.

Contrary to the situation for C, Si exhibits high solubility in all Fe-baring intermetallics, similarly to Al. Moreover, the energy of solution is often negative and large, which suggests that Si additions may also increase the stability of FeBe$_2$, FeBe$_5$ and (Fe,Al)Be$_2$. Note that the accommodation mechanism in all three phases is substitutional (on Be sites).

Whilst elemental silicon precipitates have been observed in some Be alloys~\cite{Rooksby1962a}, the current results suggest that the presence of Fe-baring intermetallics will provide strong sinks for Si dissolution, and therefore act as Si scavengers if a suitable heat treatment is applied. This is in agreement with observations by Rooksby and Green~\cite{Rooksby1962a} that in some Fe-containing Be alloy samples, Si was detected by chemical analysis but no elemental Si was found via XRD analysis.
Scavenging of Si impurities by careful alloying additions of Fe may then be beneficial for the mechanical properties of the alloy~\cite{Burr2015a}.

\subsection{Hydrogen and helium}
The main source of He in Be alloys is through exposure to $\alpha$ radiation or neutron radiation via the following nuclear reactions \cite{ENDF/B-VII.1}:
\begin{equation}
\label{eq:Be1_tab}
\begin{tabular}{*{8}{p{1.35cm}}}
\multicolumn{3}{l}{ $ {^9\text{Be}} + n \rightarrow {^6\text{He}} + {^4\text{He}} $ } \\
&\multicolumn{3}{l}{ $ {^6\text{He}} \rightarrow {^6\text{Li}} + \beta^- $ } \\
\multicolumn{2}{p{2.5cm}}{} & \multicolumn{3}{l}{ $ {^6\text{Li}} + n \rightarrow {^4\text{He}} + {^3\text{H}} $ }
\end{tabular}
\end{equation}
\begin{equation}
\label{eq:Be2_tab}
\begin{tabular}{*{8}{p{1.35cm}}}
\multicolumn{3}{l}{ $ {^9\text{Be}} + n \rightarrow {^8\text{Be}} + 2n $ } \\
&\multicolumn{3}{l}{ $ {^8\text{Be}} \rightarrow 2 {~^4\text{He}} $ }
\end{tabular}
\end{equation}
both of which yield two He atoms. Hydrogen isotopes, on the other hand may come from a variety of sources depending on the local environment, the most likely are surface corrosion of water, proton bombardment, neutron decay (free neutrons have a half-life of \SI{881.5}{s}) or via reaction~\ref{eq:Be1_tab}.

The results in Table~\ref{tab:extrinsic} are consistent with the very low solubility of H in Be metal, as predicted by previous studies \cite{Krimmel1996,Ganchenkova2009,Allouche2010a,Zhang2012}. While the solution energies are less unfavourable in Fe-baring intermetallics they are still  positive (\SIrange{0.4}{0.8}{eV}). There is, however, a favourable reaction energy for H to form BeH$_2$ following,
\begin{equation}
\text{Be}_{(s)} + \text{H}_{2(g)}  \xrightarrow{\SI{-0.17}{eV}}  \text{Be}\text{H}_{2(s)} \\
\end{equation}
if H$_2$ gas is present, or
\begin{equation}
\text{Be}_{(s)} + 2 \text{H}_i  \xrightarrow{\SI{-3.87}{eV}}  \text{Be}\text{H}_{2(s)} \\
\end{equation}
if H atoms are coming from interstitial sites in the bulk metal, as may be the case for Be components used in fusion and space applications.

Previous work showed that H is a fast diffusing species in Be metal with an activation energy for migration of \SI{0.4}{eV}, high diffusion anisotropy and strong sensitivity to traps \cite{Allouche2010a,Middleburgh2011}. Quantum tunnelling effects may further increase the mobility of H atoms in Be \cite{Flynn1970}. This suggests that despite the high stoichiometric ratio of H to Be, the hydride may form in Be alloys with dilute levels of H.

Regarding He, the results in Table~\ref{tab:extrinsic} indicate that it is highly insoluble in all phases considered, consistent with it being an inert gas. Furthermore, solution energies are all above 3 eV and consistently higher than equivalent values for H. This suggests that He bubbles are expected to form if Be alloys are exposed to $\alpha$ or neutron radiation, as observed experimentally \cite{Klimenkov2013,Barnes1959}. In turn, the He bubbles will cause changes in dimensional (swelling) and mechanical properties of the system, which are undesirable for the long term use of Be alloys in a radiation environment. 

\subsection{Oxygen}
Oxygen has a strong affinity to metals due to bond formation, hence the highly favourable formation energy of BeO (\SI{-6.06}{eV}) and largely negative  solution energies reported in Table~\ref{tab:extrinsic}.  Notably none of the intermetallics accommodate O more favourably than Be metal.
FeBe$_2$ in particular has a much lower oxygen affinity, though the solution energy of O at an interstitial site is still negative. 
Whilst this provides incomplete information regarding the oxidation of these phases --- as kinetics of diffusion or reaction may dominate the oxidation process --- it suggests there is a larger driving force for the incorporation of oxygen in the parent Be metal than the intermetallic phases.
Previous studies also indicate that O exhibits a large migration barrier of \SI{1.63}{eV} in bulk Be metal \cite{Middleburgh2011,Zhang2012a}, suggesting that the ageing kinetics of Be alloys via BeO formation are slow (though grain boundary and surface diffusion of O were not investigated).

\section{Conclusions}
\label{sec:con}

Based on calculations of thermodynamics, this study suggests that Fe additions, combined with suitable heat treatments, may have a strong influence on the equilibrium distribution of some minor components or impurity elements especially Al and Si.  In particular, Al and Si exhibit a strong driving force for incorporation into Fe containing intermetallics; a process that may be in competition with the segregation of Al and Si to grain boundaries of Be metal. Depending on to the kinetic effect of diffusion and segregation (not investigated here), the addition of Fe may be beneficial for the mechanical properties of Be alloys by limiting grain boundary embrittlement caused by Si and Al.

Conversely to Al and Si, it is not favourable for Li, a potential activation product of Be, to be incorporated into any of the phases studied here including the Fe-baring intermetallic phases. It may, however, form intermetallic compounds by reacting with other impurities, such as Mg, Al or H and O. 
The same holds true for Mg but while the formation enthalpy for MgBe$_{13}$ is positive it is sufficiently small that this phase may yet be shown to be stable.

Given that it is an inert gas it is not surprising that He accommodation is not favoured in any of the phases studied. Thus, He will exhibit extremely limited solubility and tend to form bubbles unless released out of the alloy. More surprisingly, H also exhibits limited solubility, although it is somewhat less unfavourable in some intermetallic phases than in Be metal. Conversely, H will form BeH$_2$ hydride if a sufficiently high concentration of H is reached, which may have a deleterious effect on tritium retention.
 
Carbon is not soluble in either Be metal or Be-Fe(-Al) intermetallics. Instead C additions will form Be$_2$C, which is a stable phase.
Finally, oxygen reacts strongly with Be to form BeO, but also exhibits a strong heat of solution in all phases studied here, especially Be metal.

\section{Acknowledgments}
Computational resources were provided by the Imperial College London HPC facility and the Australian National Computational Infrastructures: NCI-Raijin and MASSIVE. EPSRC and ANSTO are acknowledged for financial support, and thanks are extended to Lyndon Edwards for his contributions.

\section*{References}

\bibliographystyle{model1a-num-names}

\bibliography{/Users/pab07/Documents/papers/library}			








\end{document}